\documentclass[review]{elsarticle}
\journal{Journal of \LaTeX\ Templates}

\usepackage{lineno,hyperref}
\modulolinenumbers[1]
\usepackage{graphicx,epstopdf}
\usepackage{enumitem}
\usepackage{mathtools}  
\usepackage{float}
\usepackage{booktabs}
\usepackage{subcaption}
\usepackage{siunitx}
\usepackage{caption}
\usepackage{float}
\usepackage[miktex]{gnuplottex}
\usepackage{latexsym}
\usepackage{keyval}
\usepackage{ifthen}
\usepackage{moreverb}
\usepackage{pgfplots}
\usepackage{pgfplotstable}
\pgfplotsset{compat=newest}
\usepackage{tikz}

\newcommand{\numrangenorm}{\numrange[range-phrase= --]}

\begin{document}
	\begin{frontmatter}
		
		\title{LAB4D: A Low Power, Multi-GSa/s, Transient Digitizer with Sampling Timebase Trimming Capabilities}
		
		
		\author[mymainaddress]{Jarred M. Roberts\corref{mycorrespondingauthor}}
		\cortext[mycorrespondingauthor]{Corresponding author}
		\ead{jrobe8@hawaii.edu}
		\author[mymainaddress]{Gary S. Varner}
		\author[OSUaddress]{Patrick Allison}
		\author[mymainaddress]{Brendan Fox}
		\author[CHICAGOaddress]{Eric Oberla}
		\author[mymainaddress]{Ben Rotter}
		\author[HES-SO]{Stefan Spack}

		\address[mymainaddress]{Department of Physics and Astronomy, University of Hawai'i at Manoa,\\2505 Correa Rd., Honolulu, HI-96822 , USA}
		\address[OSUaddress]{Department of Physics, Ohio State University,\\191 W Woodruff Ave., Columbus, OH 43210 , USA}
		\address[CHICAGOaddress]{Department of Physics, University of Chicago,\\5720 S. Ellis Ave., Chicago, IL 60637 , USA}
		\address[HES-SO]{University of Applied Sciences of Western Switzerland, School of Engineering and Architecture Fribourg,\\ Route de Moutier 14, 2800 Delémont, Switzerland}


		\begin{abstract}
			The LAB4D is a new application-specific integrated circuit (ASIC) of the Large Analog Bandwidth Recorder and Digitizer with Ordered Readout (LABRADOR) family, for use in direct wideband radio frequency digitization such as is used in ultrahigh energy neutrino and cosmic ray astrophysics. The LAB4D is a single channel switched-capacitor array (SCA) 12-bit sampler with integrated analog-to-digital converters (ADC), developed in the TSMC \SI{0.25}{\micro\m} process. The LAB4D, operating at $3.2\,\textrm{GSa}/\textrm{s}$, contains 4096 total samples arranged in 32 windows, for a total record length of \SI{1280}{\nano\s}. The $3\,\textrm{dB}$ bandwidth is approximately \SI{1.3}{\giga\hertz}, with a directly-coupled \SI{50}{\ohm} input. This represents a factor of 16 increase in the sample depth and an increase in analog bandwidth in comparison to the previous generation (LAB3) digitizer. Individually addressable windows allow for sampling and digitization to occur simultaneously, leading to nearly dead time-free \si{\kilo\hertz} readout rates. All biases and current references are generated via internal digital-to-analog converters (DACs), resulting in a digitizer that requires minimal support circuitry. In addition, the LAB4D contains sample cell timebase trimming capabilities, reducing the intrinsic sample-to-sample time variance to less than \SI{5}{\pico\s}, an improvement of about 80\%. This feature allows the LAB4D to be used in precision timing applications and reduces post-hoc calibration requirements.
		\end{abstract}

		
	\end{frontmatter}
	

	
	\section{Introduction}

	There has been an increasing interest in the development of CMOS switched capacitor array (SCA) samplers due to their low-cost and high performance. These devices have been thoroughly written about in high energy physics literature \cite{kleinfelder_SCA_1988}\cite{kleinfelder_SCA_1990}\cite{Kleinfelder:1992nda}\cite{lee_CMOS_1991}\cite{Haller_SCA_1994}\cite{Haller_SCA_1994_2}. Many have accomplished digitization speeds high enough for greater than Nyquist sampling of a GHz analog bandwidth signal. In addition to their use in precision photon timing \cite{Brönnimann_1999}\cite{Ritt_2004}, these $\geq$GSa/s devices are being used in experiments designed to detect neutrinos \cite{IceCube}\cite{IceCube_2013}\cite{Kleinfelder_Neutrinos_2003}  and gamma-rays \cite{Delagnes_2006}. SCA samplers are cost-effective alternatives to analog-to-digital converters (ADCs) due to their excellent timing, recording, and high resolution amplitude \cite{Genat_2009}\cite{Breton_2011}. 
	
	The analog nature of SCA samplers also limits their widespread acceptance, requiring additional resources and time consuming post-hoc calibrations in order to achieve high performance. To simplify the integration and usage of these digitizers by minimizing or sometimes eliminating the need for on-line or off-line calibrations via software, the LAB4 ASIC design was developed with the goal to demonstrate the ability to trim the inherently non-uniform timebase generated by the CMOS voltage-controlled delay line (VCDL) by the ASIC itself, using simple internal digital-to-analog converters (DACs). A further simplification involved moving all previously-external voltage biases and current references to internal DACs, resulting in the significant reduction of support electronics needed for digitization.

	\begin{figure}[t]
		\centerline{\includegraphics[width=0.6\textwidth]{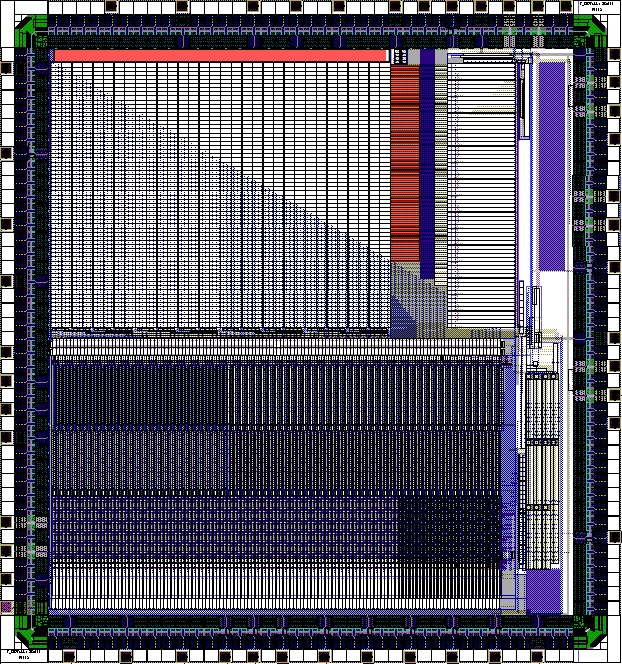}}
		\caption{Overview of the 4th-generation LABRADOR ASIC, revision D (LAB4D). The lower region of the ASIC shown in this figure shows the primary sample cells. Directly above the primary sample cells are the secondary sample arrays. Above the secondary sample arrays is the main storage array and to the right of the storage array is a region which houses all of the control DACs.}
		\label{fig:LAB4D_die_image}
	\end{figure}
	
	Another limitation of SCA samplers is the long readout time. While data can be acquired at $\textrm{GSa}/\textrm{s}$ rates, data readout typically is significantly slower (tens of $\textrm{MSa}/\textrm{s}$). At slower rates ($\sim$\si{\kilo\hertz}), a high dead time with each readout can be linked to the lack of sufficient derandomization, and the LABRADOR family of ASICs targeted this issue. For example, a $1024\,\textrm{sample}$ readout at $10\,\textrm{MSa}/\textrm{s}$ would result in the digitizer being unavailable for \SI{102.4}{\micro\s}, which, at a \SI{1}{\kilo\hertz} trigger rate would result in 10\% dead time. The LAB4 series integrates improvements stemming from the buffered LABRADOR (BLAB) series of ASICs \cite{BLAB1} to achieve simultaneous sampling and readout; acting as a 4-event derandomizer by dividing the total number of samples into multiple windows that can be written to or read out separately. This allows a significant reduction in dead time. Assuming a random trigger probability, the LAB4D would experience only 0.1\% dead time under the previous example.

	The LAB4D, the fourth revision of the LAB4 design, was fabricated in the TSMC \SI{0.25}{\micro\m} CMOS (LO) process, and packaged in a 48-pin quad-flat no-leads (QFN) package to reduce parasitic bondwire inductance. A die image of the LAB4D ASIC is shown in Figure~\ref{fig:LAB4D_die_image}. The design and performance results of the LAB4D will be discussed.

	\section{Architecture}
	
	A variety of different CMOS switched capacitor array architectures that are similar to the LAB4D have been discussed in the literature \cite{Panebianco_1999}. A compact, minimal storage array was used in order to limit the parasitic and storage capacitance of the SCA \cite{Varner_2003}. The decision to pursue a compact storage matrix architecture was symbiotic with similar designs being explored for Monolithic Active Pixel Sensors (MAPS) for charged particles \cite{Varner_NIMA_2005}\cite{Varner_NIMA_2006}.
	
	The LAB4D architecture is a descendant of the LABRADOR-3 (LAB3) design \cite{LABRADOR_2007}, developed for the Antarctic Impulsive Transient Antenna (ANITA), an ultra-high energy neutrino and cosmic ray balloon-borne observatory. The ANITA experiment requires $\sim$100 sampling channels over \numrangenorm{200}{1200}\si{\mega\hertz} \cite{Varner_2006}. Commercial flash ADCs were impractical due to cost and power limitations. The LAB3 ASIC has been successfully deployed on four ANITA long-duration balloon flights and the LAB4D is scheduled to be deployed on future missions.
	
	The features of the LAB4D and LAB3 are summarized in Table \ref{tab:1}. All of the LABRADOR ASICs have been fabricated in the TSMC \SI{0.25}{\micro\m} CMOS (LO) process and all previous generations were packaged in a 64-pin plastic thin quad-flat package (TQFP). The LAB4D was the first to utilize a 48-pin QFN package in order to reduce lead inductance and for improved compactness. 
	
\begin{figure}[t!]
		\begin{minipage}[b]{1.0\linewidth}\centerline
			{\includegraphics[width=1\textwidth]{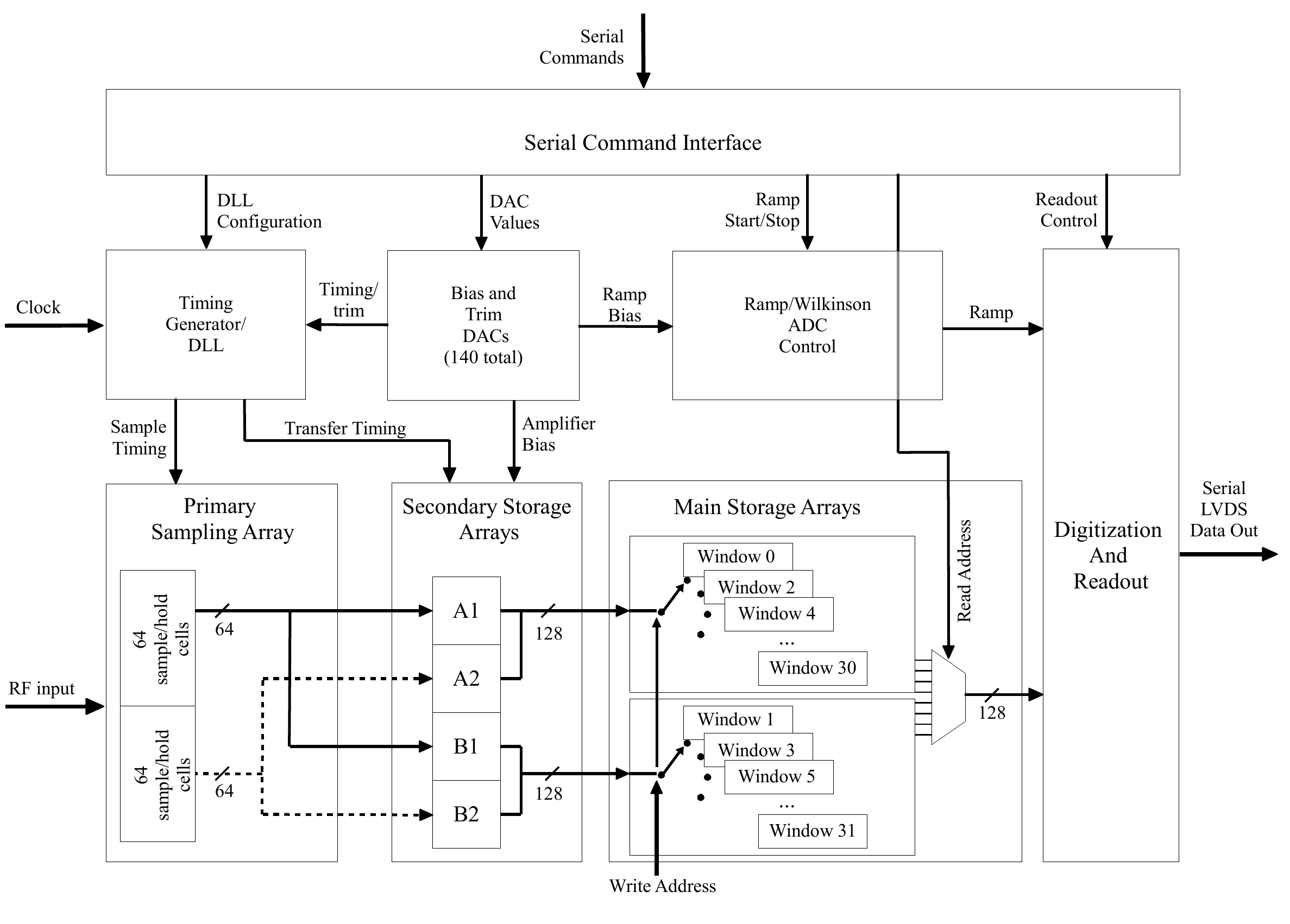}}
			\caption{Simplified block diagram for the LAB4D. RF signals are sampled in the primary array, transferred to a secondary storage array in 64-sample chunks, and finally to a 128-sample window in the main storage array of 32 windows. Secondary storage alternates between even (A1/A2) and odd (B1/B2) windows depending on the clock cycle. Timing for sampling and transfer is controlled via configurable taps from a DLL driven by the master clock, and controlled and trimmed via internal DACs. Digitization is performed via an on-die Wilkinson ADC, with
			programmable ramp current and timing. Data is then read out via a
			serial LVDS output. DLL configuration, digitization control, and DAC
			values are all programmable via a custom serial interface.}
			\label{fig:LAB4D_Data_Flow}
		\end{minipage}
	\end{figure}

	The LABRADOR ASIC ADCs feature a usable signal voltage range of \numrangenorm{0}{2.5}\si{\volt} and the LAB4D's ADC has an effective readout rate of $>$100 kSa/s. The LAB4Ds were designed as single channel RF digitizers in order to remove bondwire and on-die coupling that was present in the LAB3. A general overview of the LAB4D signal processing is shown in Figure~\ref{fig:LAB4D_Data_Flow}. As this figure shows, The LAB4D consists of a ``primary'' sampling array arranged as two blocks of 64 sample-and-hold cells which are directly connected to an externally-terminated \SI{50}{\ohm} RF input. These blocks are transferred to one of four 64-sample ``intermediate'' storage arrays. Two of the intermediate storage arrays (128 total samples each) are then transferred to the main storage array which is organized into 32 ``windows'', corresponding to 4096 total samples. See section~\ref{sec:sampling_array} for more details regarding the sampling architecture.

	
	\begin {table}[t!] 
	\caption {Summary of LABRADOR Specifications and Improvements}
	\begin{center}
		\begin{tabular}{@{} *5l @{}}  \toprule
			\emph{Item} & \emph{LAB3} & \emph{LAB4D} \\\midrule
			$\#$ RF Inputs    & 9 & 1    \\ 
			Sample-and-hold cells per input & 260 & 128\\ 
			Total number of samples & 2340 & 4096\\
			$\#$ of ADCs & 2340 & 128\\
			Sampling speed control & Analog & Delay-locked loop\\
			RMS of sampling intervals & 50 & 2.5 \\
			3\hspace{0.05cm}dB bandwidth for 50 ohm input (MHz) & \si{900} & \si{1300}\\
			Sampling speed [GSa/s] & 2.6 & 3.2\footnotemark{}\\
			Package & 128-pin QFP & 48-pin QFN \\

			\bottomrule
			\hline
		\end{tabular}
	\end{center}
	\label{tab:1}
	\end {table}

	\footnotetext{This is a desired operating point and not a design limitation}

	The 128 sampling intervals run in parallel with a voltage-controlled delay line VCDL and delay-locked-loop (DLL). The DLL fixes the total delay to a master clock that is externally-supplied. Using an external on-board \SI{25}{\mega\hertz} clock, the sampling rate is 3.2 GSa/s. A unique feature of the LAB4D is the ability to tune the individual sampling intervals; allowing the mitigation of the sampling differential non-uniformity in the hardware using the built-in trim DACs. Additional internal DACs are used to provide voltage biases for the analog transfer and digitization, which allows for the optimization of power consumption. Further details on the timing generation of the LAB4D will be discussed in section~\ref{sec:timing_generation}.

	The analog bandwidth of the LAB4D has been expanded in the high frequency region by \SI{400}{\mega\hertz} in comparison to previous generations of the ASIC. The LAB4D has a storage array of 4096 cells, arranged in 32 blocks of 128. For the ANITA mission, 8 of these blocks will be read out for each event, which at 3.2 GSa/s results in \SI{320}{\nano\s} duration waveforms for each event; a 100\% increase in comparison to the LAB3. Windows in the main storage array can be written to and read from in a random-access manner using an externally-provided write and read address. The window address is selected for readout by an independent RF trigger unit within the ANITA instrument crate. The storage bank acts as a circular buffer, where the write pointer skips the banks reserved for readout so that they are not overwritten until the readout is complete. The selected window is then digitized in parallel via a Wilkinson (ramp) ADC, and then the final digitized data is shifted out using a high-speed LVDS output. 
	
	The use of Gray code was also implemented within the reference counter of the Wilkinson ADC converter. The counter is incremented on every rising edge of the externally supplied Wilkinson clock and is distributed to each of the 128 conversion registers. The use of Gray code in the conversion register prevents erroneous outputs that are intrinsic to standard binary counters when there are very small delays between counter bits. Readout speeds of \SI{20}{\micro\s} per window are practical, resulting in a readout speed of $6.4\,\textrm{MSa/s}$.


	\subsection{Sampling Array}
	\label{sec:sampling_array}

\begin{figure}[h!]
		\centering
		\begin{minipage}[b]{1\linewidth}
			\hspace{0.63cm}
			{\includegraphics[height=5.4cm]{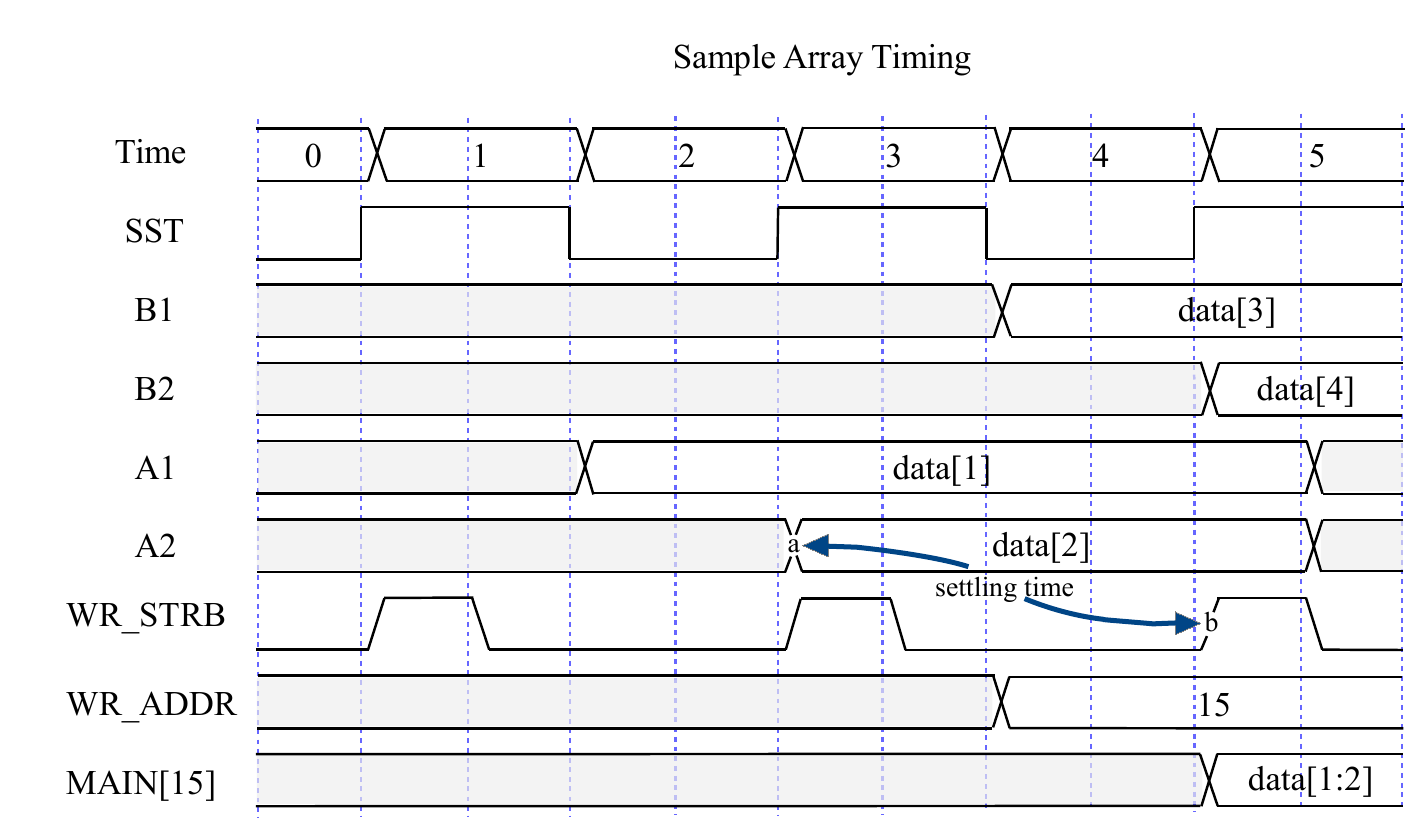}}
			\vspace{0.5cm}
		\end{minipage}
		\centering
		\begin{minipage}[b]{1\linewidth}
			\centering
			{\includegraphics[height=5.8cm]{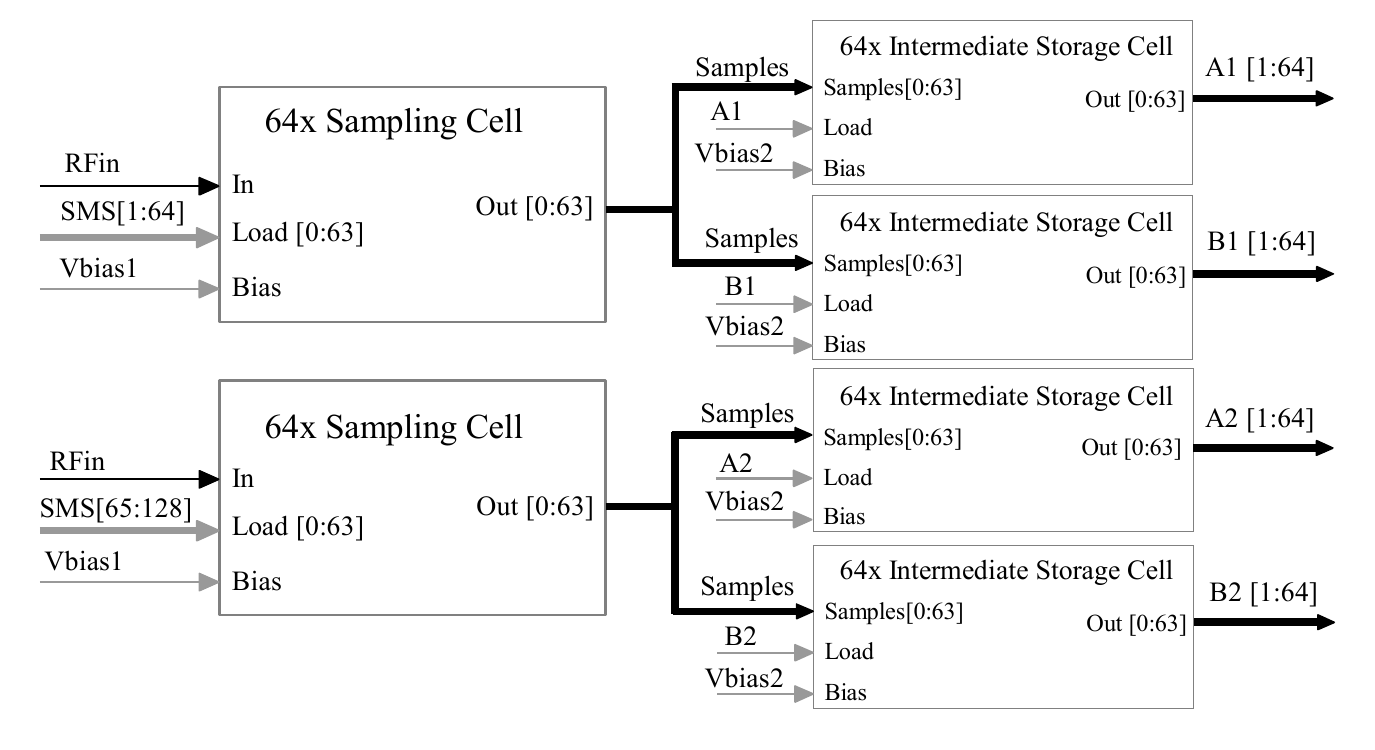}}
		\end{minipage}
		\caption{
			\textbf{Upper}: Timing diagram for the two-stage transfer in the LAB4D. Each time period in the timing diagram corresponds to 64 samples. 
			\textbf{Lower}: Data flow in the LAB4D sampling array. Samples are first acquired in the 2 chunks of 64 sampling
			cells, with timing determined by the SMS[1:128] signals from the VCDL. Vbias/Vbias2 are bias voltages which determine
			the output amplifier drive strength.}
		\label{fig:Sample_Array}
	\end{figure}

	As shown in Figure~\ref{fig:LAB4D_Data_Flow}, the LAB4D features a unique two-stage sample transfer architecture which allows for an effective doubling of the settling time into the main storage array. In short, this sampling array utilizes what we call a \textit{ping-pong} intermediate storage cell method where two sets of 64 primary sampling cells are paired with two intermediate storage cells; essentially working as a 4-event derandomizer. Data from the primary sampling cells are transferred in 64-sample chunks to the intermediate storage arrays (labeled A1, A2, B1, and B2). When the 128 total samples of intermediate storage cell 1 (A1 and B1) become ``full", intermediate storage cell 2 (A2 and B2) is then written to; allowing cell 1 to be transferred to an address (WR\_ADDR) in the main storage array (labeled ``15'' in Figure~\ref{fig:Sample_Array}) in a subsequent sampling cycle when the write strobe (WR\_STRB) is active.

	It is the alternating between which sampling array and intermediate storage array the sampling cell writes to, that resembles a ping-pong ball constantly changing its side of the table and, hence, the origin of the terminology. The two-stage transfer that is shown, extends the settling time for the secondary array to a full sampling cycle. The result is an improved decoupling of the primary array from the sampling array, which is necessary given the fact that the main storage array occupies most of the physical space of the ASIC (see Figure~\ref{fig:LAB4D_die_image}) and the samples that are being stored must travel long routes across the chip to reach the designated storage cell.


	\subsection{Timing Generation}
	\label{sec:timing_generation}

\begin{figure}[t!]		
		\centering
		\begin{minipage}[h]{1\linewidth}
			\centering
			{\includegraphics[width=12cm]{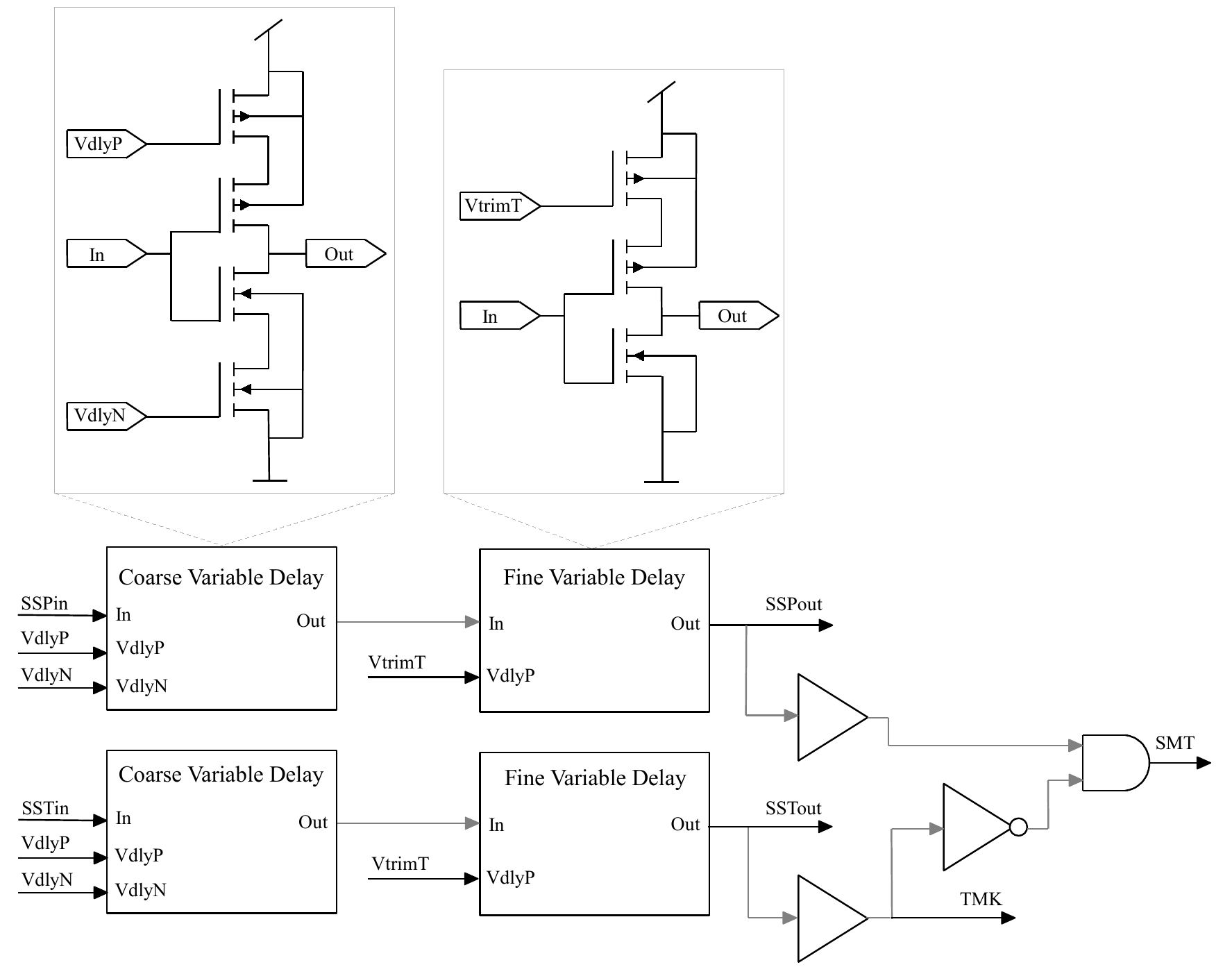}}
		\end{minipage}
		\caption{
			\textbf{Upper}: Simplified schematic diagrams of the variable delay elements. VdlyP/VdlyN are common controls for the entire delay chain, whereas VtrimT is per-delay. \textbf{Lower}: Simplified block diagram of a LAB4D delay block, one of 128 present in the delay chain for the main sampling clock, SST. At each delay, the output goes to a timing generator (TMK) which provide the programmable timing strobes for internal controls, one of which is the start strobe (SSP). Samples are tracked when SMT is active, formed by the combination of SSP and not SST.}
		
		\label{fig:timing_generation}
	\end{figure}

	As shown in figure~\ref{fig:timing_generation}, timing for the 128 primary sample-and-hold cells in the LAB4D is controlled by taps from the VCDL in the timing generator. The VCDL is a 128-element delay line, with each element consisting of 2 voltage controlled delays, implemented as current-starved inverters.

	The total delay of the VCDL is locked to an external clock using a DLL which compares the external clock after propagating through the VCDL to a non-delayed version of the same clock. The phase comparator output then drives a charge pump, which adjusts the VdlyN control voltage, that in turn adjusts the propagation speed of the high-to-low transitions of the first stage of current starved inverter. This signal is common to all 128 delay stages.  
	
	Individual delay stage adjustment is done by adjusting separate VtrimT, set via individual DACs for all 128 stages. By trimming these DAC values, the intrinsic differences between current-starved delays due to process variations can be nulled.  Identical copies of these delay pairs are used to delay the priming signal (SSP) and the timing signal (SST), which are used to synthesize the actual Track/Hold signal (SMT). SSP precedes SST and sets the SMT signal high, putting this delay stage's Switched Capacitor into Track mode. The later arrival of SST, which is a direct and delayed copy of the SSTin reference clock, drives SMT into Hold mode. At each stage a copy of the delayed SST edge (TMK) is extracted and used by a programmable timing generator to synthesize required internal timing strobes, such as SSP (which is actually started at the end of the propagation of the previous SSTin cycle, to permit it to be present prior to the arrival of SST).

	The primary purpose of the DLL is to compensate for intrinsic changes in current-starved inverter propagation due to changes in temperature. VdlyP does not adjust the critical rising or timing edge of SST, but rather the falling edge. It needs only be set to a reasonable value, via DAC, such that the 50\% duty cycle of the incoming SSTin clock is not made excessively asymmetric.


	\subsection{Implementation}
	
	\begin{figure}[t!]	
		\centering
		\begin{minipage}[t]{0.7\linewidth}\centerline{\includegraphics[height=8.5cm]{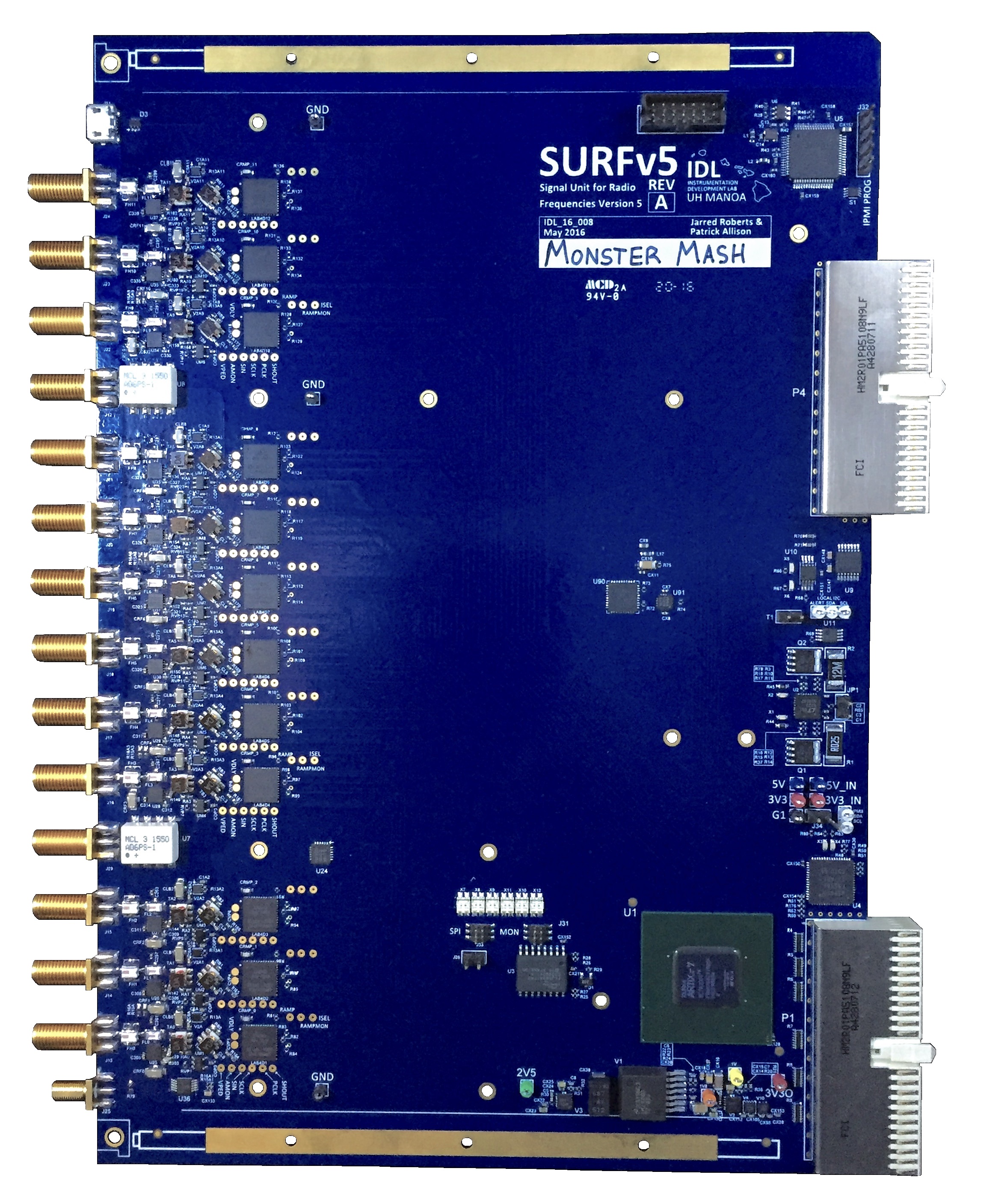}}		
		\end{minipage}	
		\caption{Image of the Sampling Unit for Radio Frequencies, version 5 (SURFv5), used for LAB4D characterization.
			Each SURFv5 contains 12 LAB4D ASICs, packaged in a 48-pin QFN package. The LAB4Ds are visible
			on the left section of the board immediately after the RF conditioning components. The printed circuit board dimensions are 6.3 x 9.2 inches.}
		\label{fig:Monster_Mash_edit}	
	\end{figure}
	
	Characterization of the LAB4D performance was performed using a 12-channel digitizer designed for the ANITA experiment,
	the \textit{Sampling Unit for Radio Frequencies}, version 5 (SURFv5). The SURFv5 is a CompactPCI-compatible 6U printed circuit board (PCB), with a single Xilinx Artix-7 field-programmable gate array (FPGA) interfacing with all LAB4Ds and providing a CPU interface for data readout and LAB4D configuration. The incoming RF signals are conditioned with a \numrangenorm{200}{1200}\si{\mega\hertz} bandpass using Mini-Circuits HFCV-145+ low-pass and LFCN-1200+ high-pass filters. In addition, a copy of the incoming signal was coupled off using a Mini-Circuits TCD-13-4X+ for RF power monitoring, and a calibration signal common to all LAB4Ds was coupled in using an identical coupler. Finally, the AC-coupled signal was \SI{50}{\ohm} terminated to an adjustable DC voltage.	A total of nine SURFv5s were manufactured, and two SURFv5s have currently been tested and characterized, for a total of 24 LAB4Ds. Performance of the 24 LAB4Ds were broadly comparable.  Each SURFv5 was assigned an identifier associated with a popular Oahu surfing location, as shown in Figure~\ref{fig:Monster_Mash_edit}.


	\section{Results}
	
	The performance of the LAB4D ASIC as implemented on the SURFv5 was characterized to determine its suitability for RF
	digitization. Specifically, the noise level, linearity, working range, sample-to-sample timing variation, and analog
	bandwidth were measured. Next, the stability of these measurements with respect to operating temperature was characterized.
	
	Temperature variations were investigated using 3 LAB4Ds which were coupled using thermal paste to a copper heat sink
	with a Peltier heater/cooler with a separate heat sink on the opposite side, as well as a thermometer for monitoring the
	LAB4D temperature. Current through the Peltier device was varied to control the LAB4D temperature.
	
	\subsection{Noise}
	\label{sec:Noise}

\begin{figure}[t]
		\centering
		\begin{minipage}[t]{0.48\linewidth}\centerline{\includegraphics[height=4.75cm]{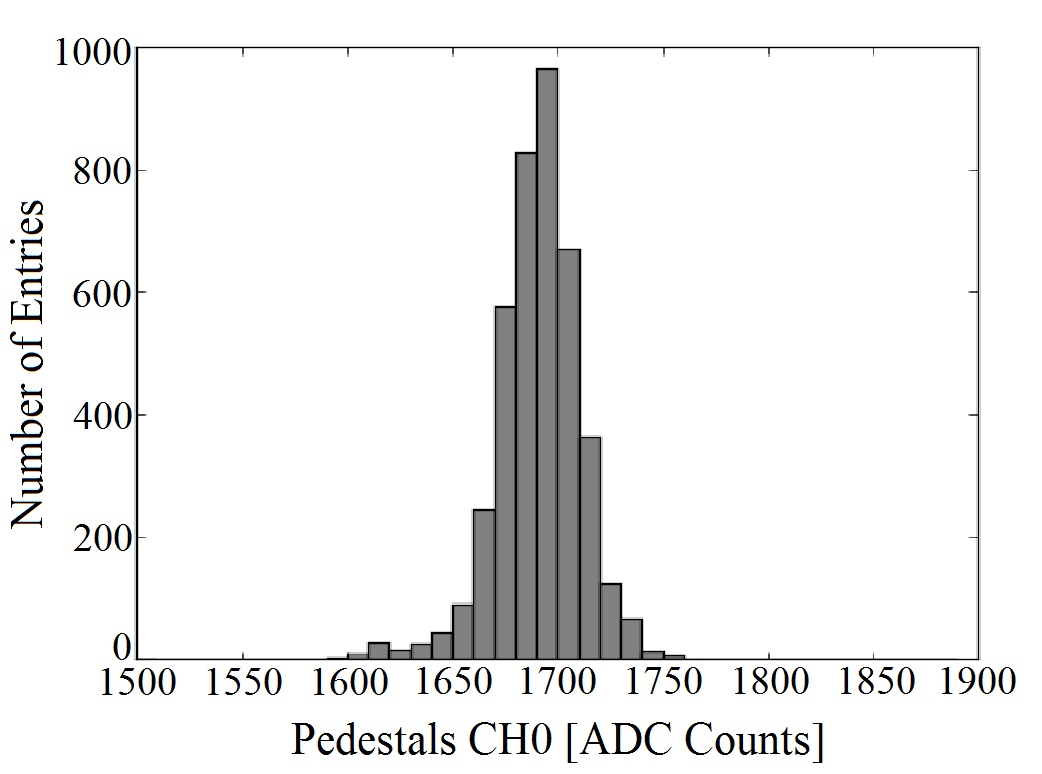}}
		\end{minipage}
		\hspace{0.02\linewidth}
		\begin{minipage}[t]{0.48\linewidth}\centerline{\includegraphics[height=4.7cm]{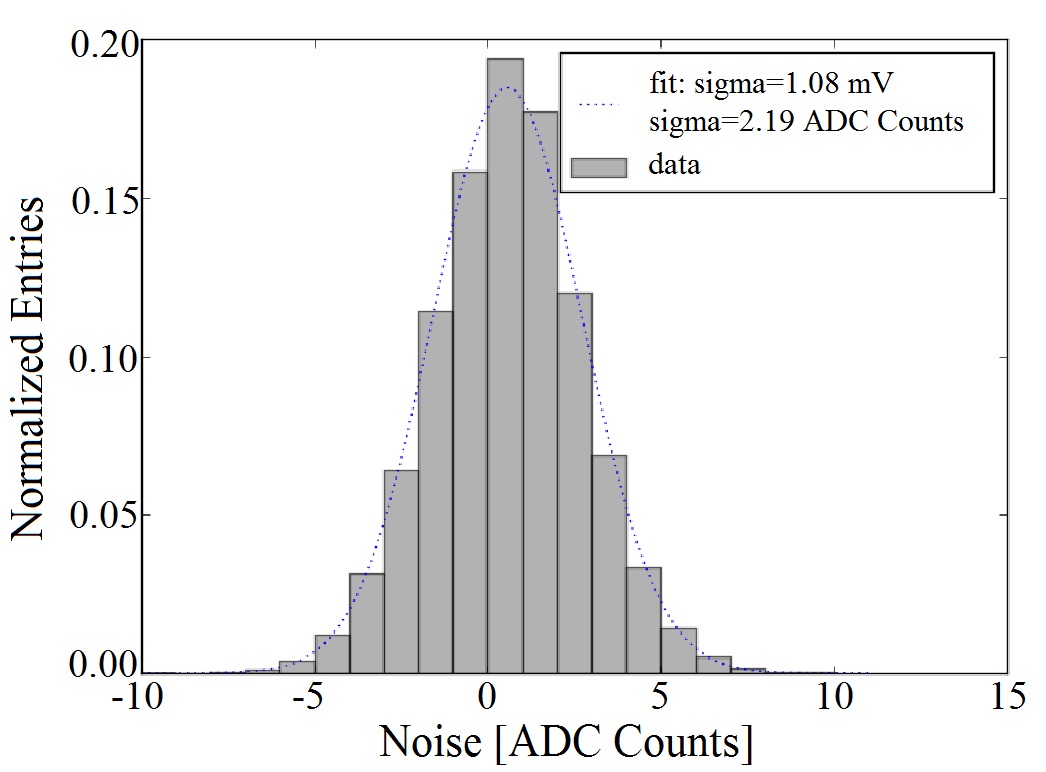}}
		\end{minipage}
		\caption{\textbf{Left}: Distribution of measured pedestal values for a typical LAB4D. The input DC voltage for this measurement was \SI{0.74}{\volt}. Measured pedestals ranged from \numrangenorm{1600}{1750} counts, which corresponds to approximately $\sim3.7\%$ of the total dynamic range. \textbf{Right}: Terminated input histogram. The spread in measured values is 2.19 ADC channels, which corresponds to \SI{1.08}{\milli\volt} using the DC transfer function in the linear region of 2 ch/mV.}
		\label{fig:Pedestals_Canoes}
	\end{figure}

	Each individual storage cell in the main storage array develops a slightly different DC offset due to non-uniformity in the
	fabrication process. These offsets, called ``pedestals'', must be measured and removed to recover the input signal. Subtraction
	of the individual pedestal can easily be done at the point when each sample is read out, since each sample's subtraction is
	independent. The measured pedestal values for all LAB4D ASICs on a single SURFv5 is shown in Figure~\ref{fig:Pedestals_Canoes}. The
	intrinsic pedestal variation results in a relatively minor $\sim3.7\%$ reduction in total dynamic range.

	Once the individual pedestals are measured and removed, a histogram of the measurements for a terminated input was obtained to determine the overall kTC noise contributions of the switched capacitor based sampling cells and the two storage cells of the LAB4D. The observed total noise is roughly saturated by these contributions and was measured to be $2.19\,\textrm{ch}$, which corresponds to approximately \SI{1}{\milli\volt} using a nominal DC transfer function of $2\,\textrm{ch}/\textrm{mV}$. This noise level is broadly consistent with other SCA samplers and previous generation LABRADOR ASICs and is not expected to be an operational
	limitation.


	\subsection{Linearity}
	
	\begin{figure}[t]
		\centering
		\begin{minipage}[h]{0.9\linewidth}\centerline{\includegraphics[width=1\textwidth]{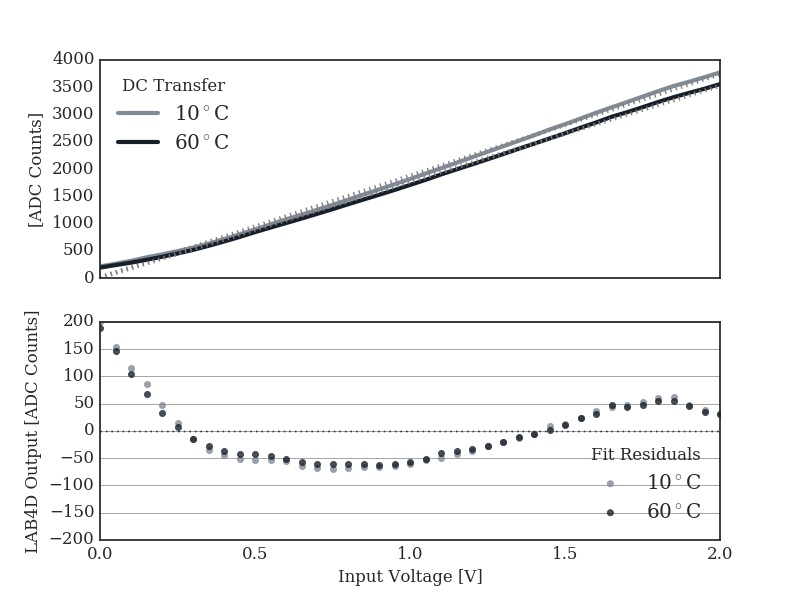}}
			\caption{A best linear fit of single-parameter DC scans of a single sample cell at multiple temperatures are shown where there is an integral non-linearity (INL) of better than 2.5\% for about 75\% of the region.}
			\label{fig:temp_thresh}
		\end{minipage}
	\end{figure} 
	
	After pedestal subtraction, the linearity of the digitization of the LAB4D was characterized. This was performed by conducting
	single-parameter scans of the DC input voltage to the LAB4D. Over a \SI{1}{\volt} span from \numrangenorm{500}{1500}~\si{\milli\volt}, the DC transfer curve shows an integral non-linearity of better than 2.5\% for most of the region, as shown in Figure~\ref{fig:temp_thresh}. Also shown in this figure, the transfer curve slope varied by approximately 0.1\%/\SI{1}{\degree}\hspace{0.05cm}C. In addition, pedestal offsets were measured to vary by approximately 0.05\%/\SI{1}{\degree}\hspace{0.05cm}C. DC scans were also conducted for 1024 LAB4D sample cells in order to determine the level of dispersion between cells. Gain dispersion within the linear region was measured to be approximately 2\% over all cells.


	\subsection{Sample-to-sample timebase variation}
	
\begin{figure}[t]
		\centering
		\begin{minipage}[t]{0.48\linewidth}\centerline{\includegraphics[height=4.5cm]{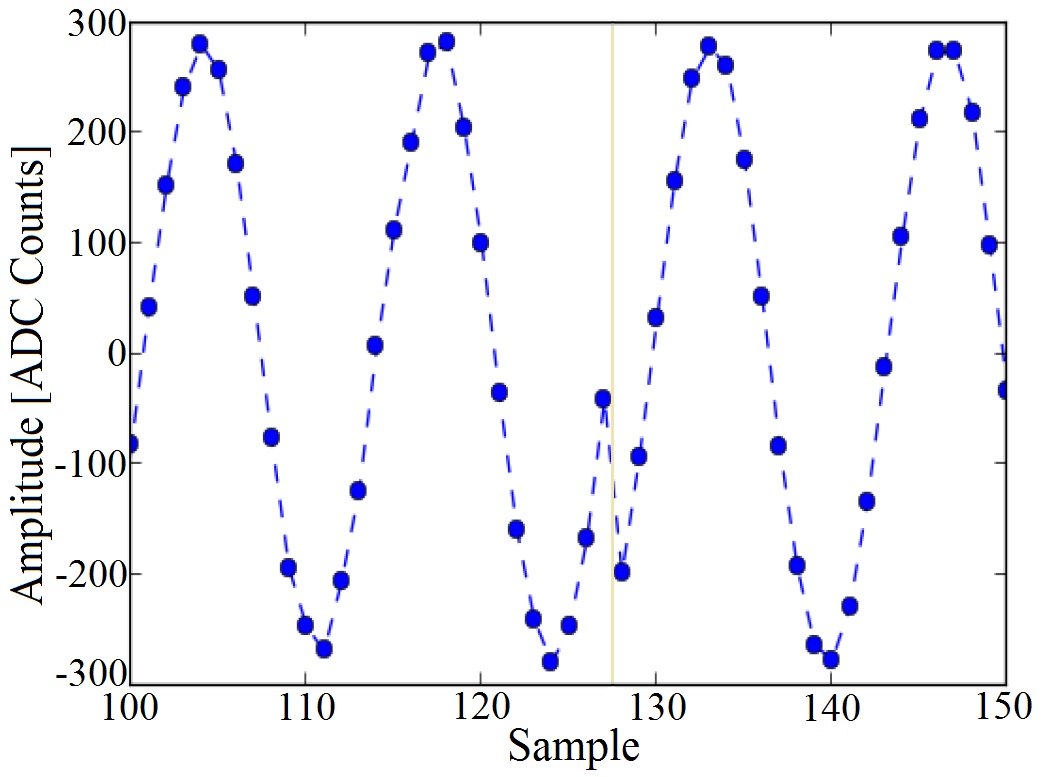}}
		\end{minipage}
		\hspace{0.02\linewidth}
		\begin{minipage}[t]{0.48\linewidth}\centerline{\includegraphics[height=4.5cm]{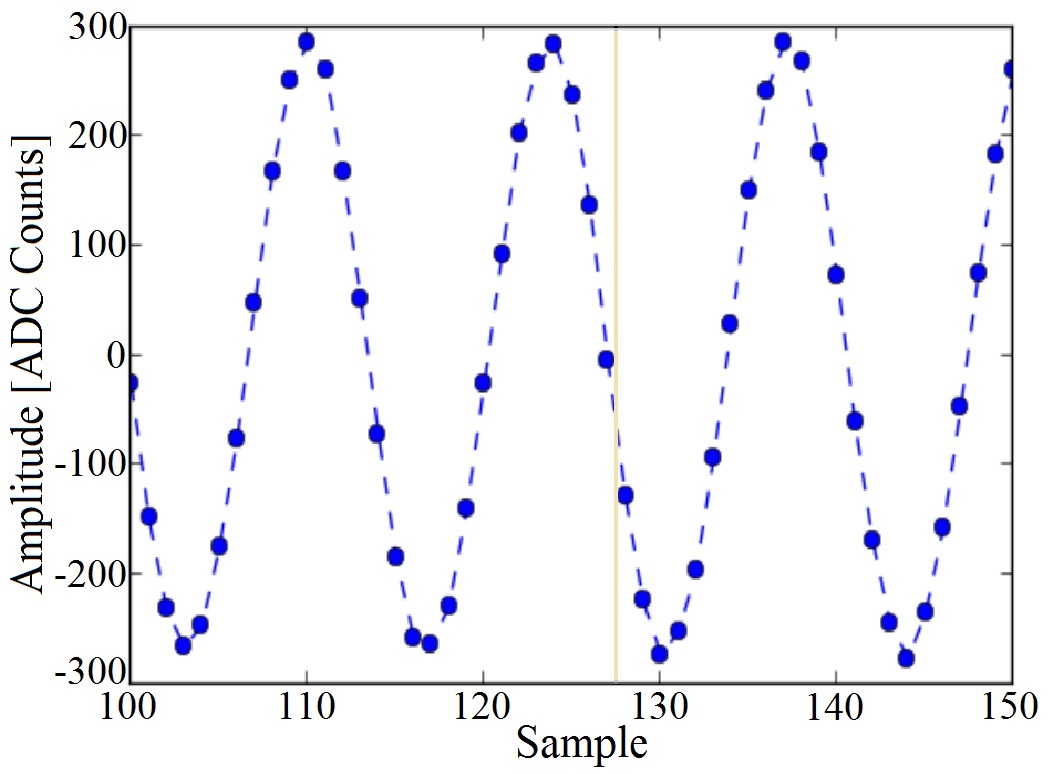}}
		\end{minipage}
		\caption{
			\textbf{Left}: Window-to-window stitching of a 235MHz sine wave before initial DLL tuning. The vertical line represents the location of the
			two window (128 sample) boundaries.
			\textbf{Right}: Window-to-window stitching of a \SI{235}{\mega\hertz} sine wave after tuning. The previously visible window boundary is now minimized.
		}
		\label{fig:DLL_tuned}
	\end{figure}

	The DLL is first optimized by sampling a \SI{235}{\mega\hertz} sine wave. The individual trim DACs are set to a common approximate
	delay, which assigns a portion of the delay for each element to be controlled by the trim delay, and the remainder to be controlled
	by the DLL. Because of internal routing in the delay line, even and odd samples are set to different initial values.
	Then, a sine wave is fit to a single window of data and VtrimT is adjusted so that the fit frequency matches
	the input frequency. An example of this optimization can be seen in Figure~\ref{fig:DLL_tuned}.  At this point, the \textit{average}
	sampling speed of the LAB4D is 128 times the external clock, and the sample-to-sample timing variations can then be tuned.

	The individual sample timing is then trimmed using the individual trim DACs. Because the DLL acts to keep the overall delay of the VCDL
	the same, adjustments in any single sample trim DAC will result in the corresponding adjustment of the timing of that sample \textit{and}
	all other samples. That is, slowing down a single sample by 1\% using the trim DACs will result in the remaining samples speeding up by
	$\frac{1}{127}\%$ to compensate.
	
	\begin{figure}[t]
		\centering
		\begin{minipage}[t]{0.48\linewidth}\centerline{\includegraphics[height=5cm]{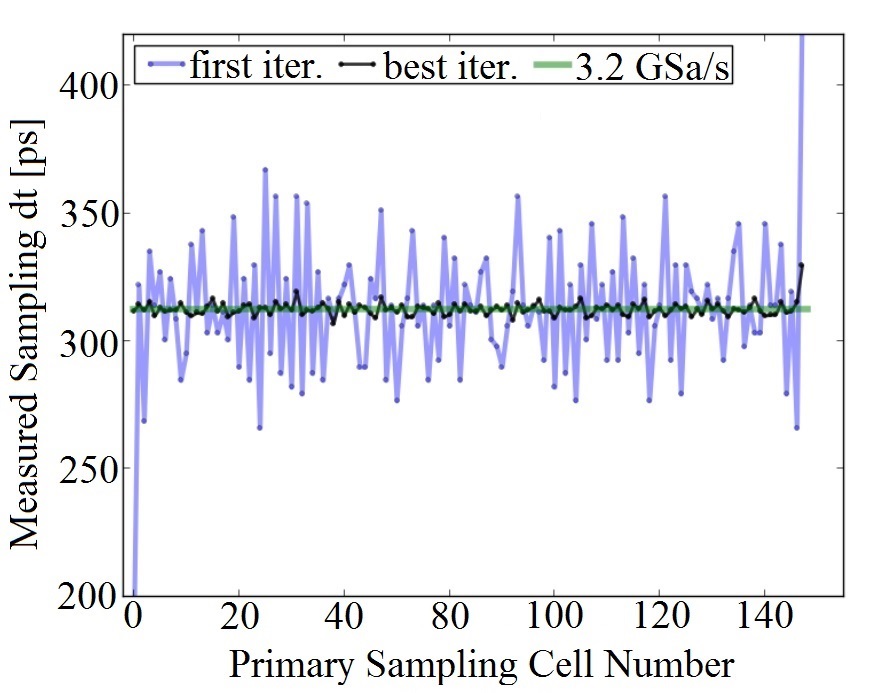}}
		\end{minipage}
		\hspace{0.02\linewidth}
		\begin{minipage}[t]{0.48\linewidth}\centerline{\includegraphics[height=5cm]{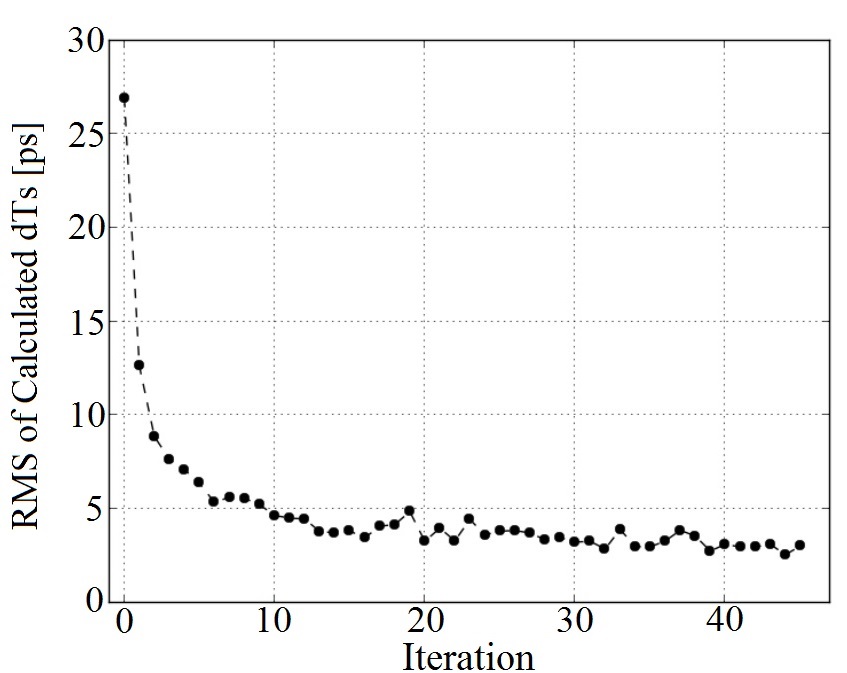}}
		\end{minipage}
		\caption{\textbf{Left}: Measured sampling intervals versus sample number before and after tuning. Routing in the LAB4D delay line results in an odd/even
			sample initial timing variation, which is eliminated after tuning. The final sample timing is currently under investigation.
			\textbf{Right}: Improvement in the RMS of the individual sample timings as a function of the number of iterations. The RMS decreases to below \SI{5}{\pico\s}
			within 10 iterations (80,000 waveforms), and further reduces to approximately \SI{3}{\pico\s} after 32 iterations.}
		\label{fig:Sampling_tuning}
	\end{figure}
	
	We therefore use an iterative minimization procedure to determine all trim DAC values simultaneously. This is done by measuring the fraction of observed samples where the observed value (relative to the subsequent sample) crosses the DC pedestal (``zero-crossing fraction''), either positive-going or negative-going, for 8000 separate waveforms. These 8000 waveforms make up a single iteration in the minimization procedures we discuss throughout this text. This fraction is a simple measure of the width of the time sample, and should be constant if the sample timing is regular. For samples that have a zero-crossing fraction greater than the average, the trim DAC value is decreased, speeding up that sample. Likewise, samples that have a lower-than-average zero-crossing fraction are slowed down by increasing the trim DAC value. Global structure in the delay line is compensated for by appropriate initial estimates of the trim DACs.

	This procedure is repeated multiple times, which progressively reduces the variation in the timebase. Within approximately 10 iterations (80,000 waveforms), the RMS variation in the sample-to-sample timing is reduced below \SI{5}{\pico\s}. The iterative procedure eventually (after about 30-40 iterations) reaches RMS timing variations of approximately \SI{3}{\pico\s}. The improvement in timing over iteration count, as well as the sample-to-sample timing, can be seen in Figure~\ref{fig:Sampling_tuning}. The decrease in improvement below \SI{5}{\pico\s} is due to the limited statistics of each iteration, rather than an intrinsic limitation of the timebase tuning procedure. The last intrawindow delay ($t_{127}-t_{126}$) contains an unknown extra delay which is currently under investigation, and results in this sample timing being a noticeable outlier from the other samples. This should be further adjustable by assigning more overall delay to the trim DACs in the calibration procedure. The current calibration procedure primarily targets the differential nonlinearity (DNL) of the timebase for simplicity as well as the fact that the largest timebase variations are an even/odd sample systematic. Future procedural improvement will focus on constraining the timebase INL as well. Initial measurements of timing variations of identical pulses in different channels on the SURFv5 currently show an \SI{8}{\pico\s} RMS variation, indicating that the INL is relatively controlled by the DLL.

	\begin{figure}[t]
		\centering
		\begin{minipage}[t]{0.48\linewidth}\centerline{\includegraphics[width=1\linewidth]{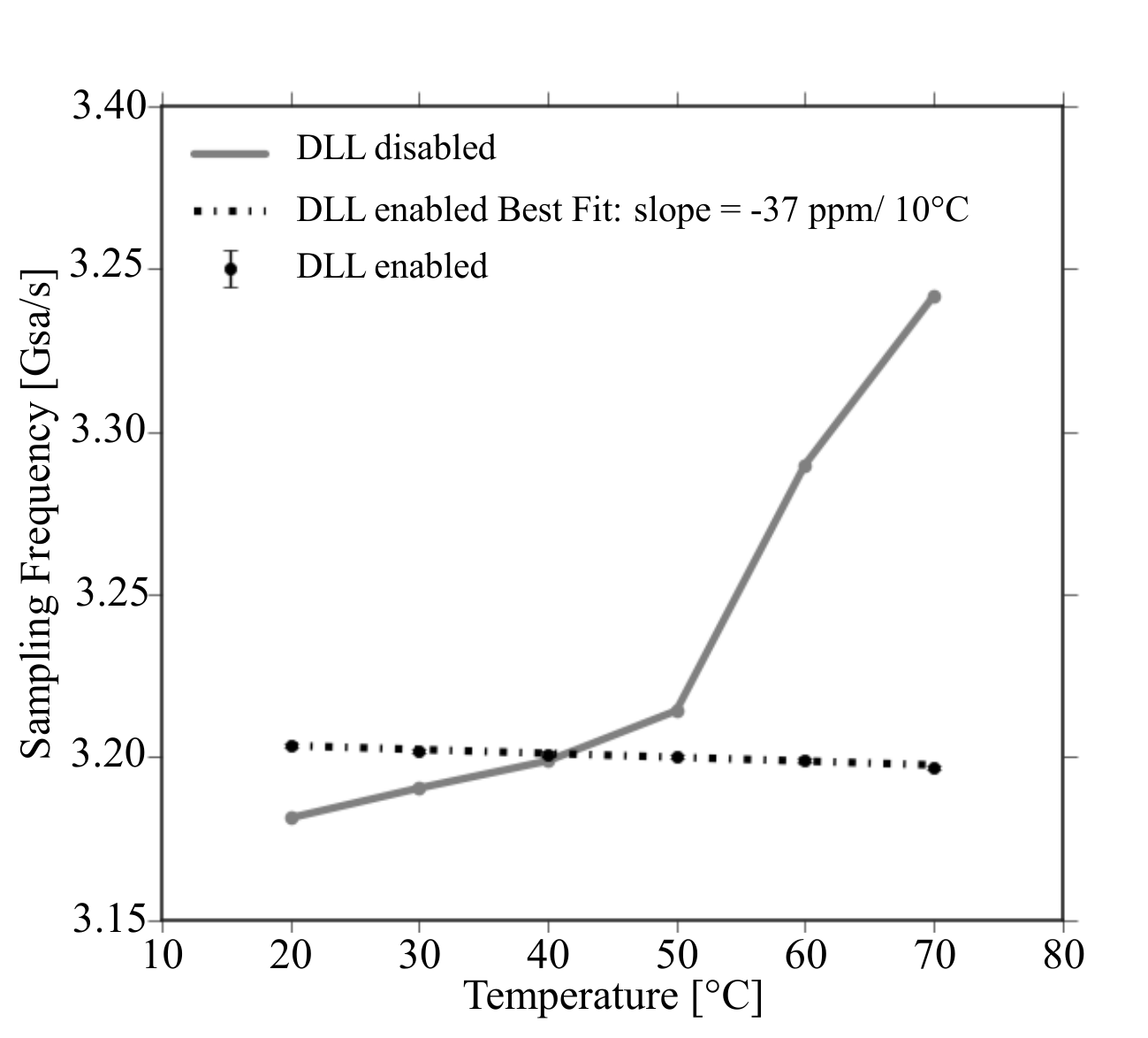}}
		\end{minipage}
		\hspace{0.00\linewidth}
		\begin{minipage}[t]{0.48\linewidth}\centerline{\includegraphics[width=1\linewidth]{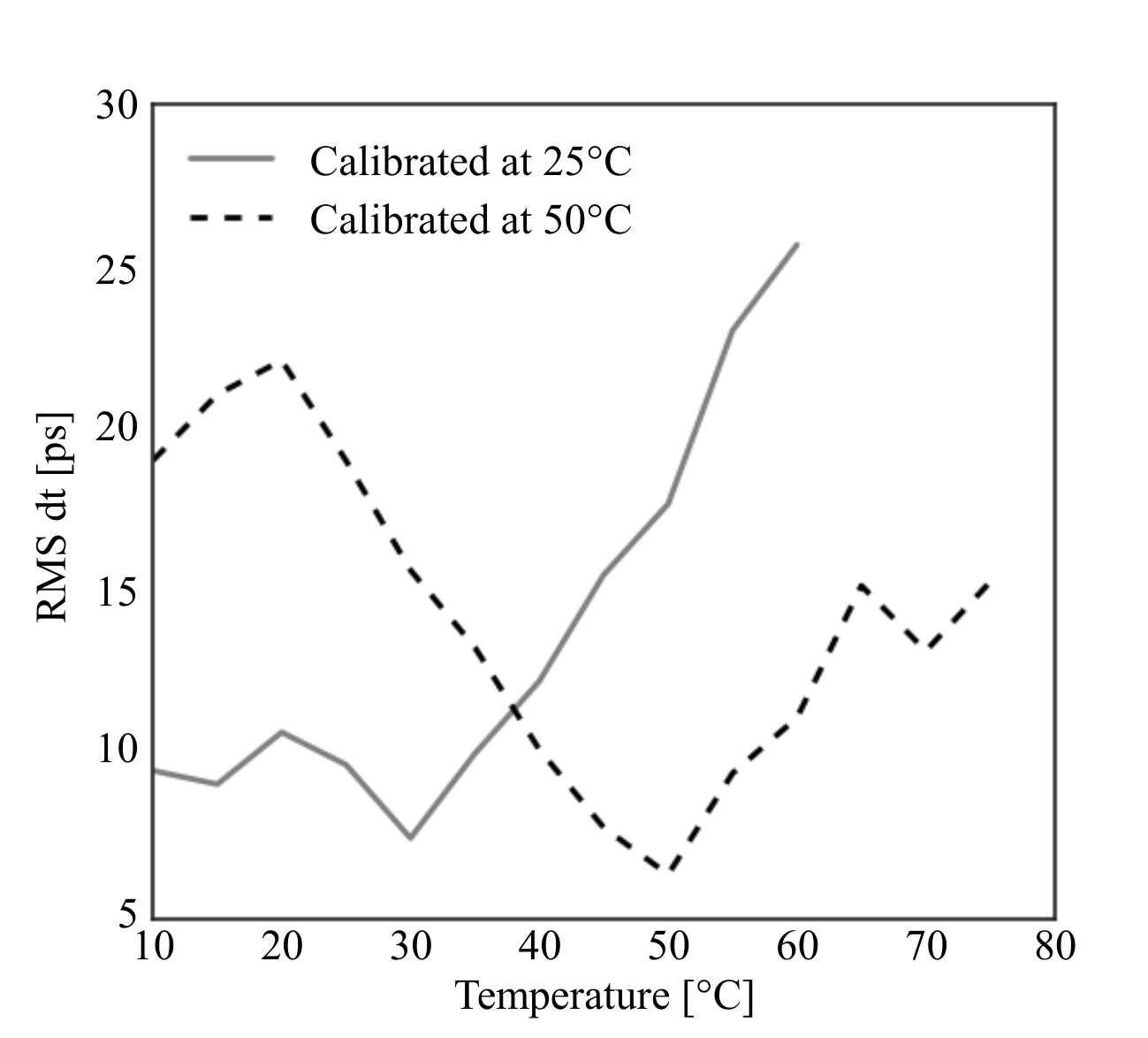}}
		\end{minipage}
		\caption{
			\textbf{Left}: Measured sampling frequency as a function of temperature both with DLL enabled (dashed line) and DLL disabled (solid line) where the 
			sampling frequency shown to be between $500-2000~\textrm{ppm}/^{\circ}\textrm{C}$, is reduced to less than $3~\textrm{ppm}/^{\circ}\textrm{C}$ by the DLL.
			\textbf{Right}: RMS variations of the sample timebase as a function of temperature when calibrated at room temperature (solid line) and \SI{50}{\degree}\hspace{0.05cm}C (dashed line), with DLL functionality always enabled.
			Sample-to-sample variability remained within a factor of 2 of the calibration within a range of $\pm10~^{\circ}\textrm{C}$.}
		\label{fig:temp_thresh_tune}
	\end{figure}

	Finally, the stability of the timebase with respect to temperature variation was also investigated. First, the DLL was tuned at a temperature of \SI{30}{\degree}\hspace{0.05cm}C and a \SI{210}{\mega\hertz} sine wave was digitized by the LAB4D. Data was then taken between \SI{10}{\degree}\hspace{0.05cm}C and \SI{60}{\degree}\hspace{0.05cm}C in \SI{10}{\degree}\hspace{0.05cm}C intervals, with the DLL enabled, and again with the DLL function
	disabled, in order to investigate the reduction in temperature sensitivity due to the DLL. The sampling rate was manually set by fixing the voltage which controls the common delay in the VCDL via an internal DAC.

	The sampling frequency of the LAB4D was determined again using the zero-crossings of the recorded data, and the sampling frequency versus temperature is shown in Figure~\ref{fig:temp_thresh_tune}. With the DLL disabled, the sampling frequency had a strong temperature dependence, between $500-2000~\textrm{ppm}/^{\circ}\textrm{C}$. The action of the DLL reduces that temperature dependence to less than $3~\textrm{ppm}/^{\circ}\textrm{C}$. 
	
	Temperature variations also affect the regularization of the sampling timebase. To explore this, the trim DACs were briefly calibrated at both \SI{25}{\degree}\hspace{0.05cm}C and \SI{50}{\degree}\hspace{0.05cm}C, and then the sample-to-sample time variation was measured from \numrangenorm{10}{75}\si{\degree}C in \SI{5}{\degree}\hspace{0.05cm}C steps. The DLL was enabled for all tests. These results are also shown in Figure~\ref{fig:temp_thresh_tune}. The sampling variations show a strong temperature dependence with temperature variance. However, over a range of $\pm10~^{\circ}\textrm{C}$ the RMS time variations remained within a factor of 2 of their original calibrations.


	\subsection{Analog bandwidth}
	
\begin{figure}[t]
		\centering
		\begin{minipage}[t]{0.48\linewidth}\centerline{\includegraphics[height=4.2cm]{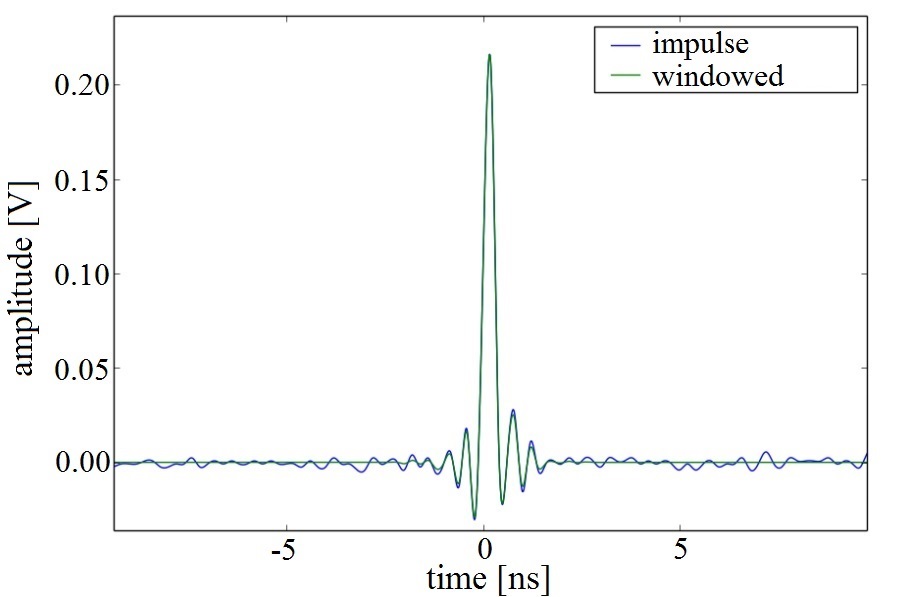}}
			\caption{Input pulse response from a Tektronix AWG5104 pulser of a Tektronix MSO 5204B pulser.}
			\label{fig:pulse_vs_windowed}
		\end{minipage}
		\hspace{0.02\linewidth}
		\begin{minipage}[t]{0.48\linewidth}\centerline{\includegraphics[height=4.3cm]{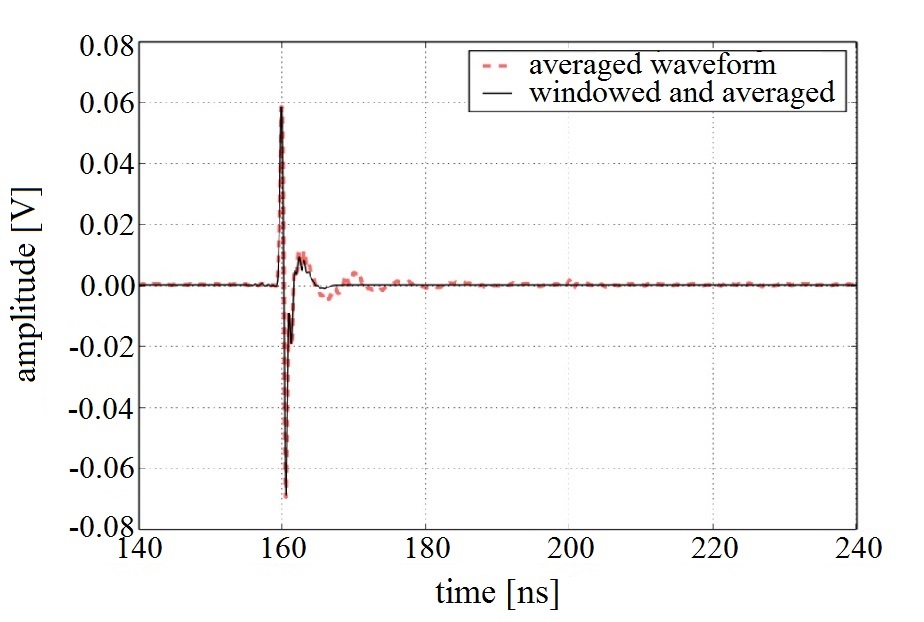}}
			\caption{LAB4D averaged and windowed averaged waveform readout.}
			\label{fig:AVG_waveform}
		\end{minipage}
	\end{figure}

	The LAB4D analog bandwidth was measured using the impulse response of
	the SURFv5. The impulse response was used to avoid effects from
	sampling persistence \cite{Oberla:2013mra} which would be present when
	using a frequency-swept continuous-wave (CW) source. The SURFv5, as was previously mentioned, has an
	RF input chain consisting of a Mini-Circuits HFCV-145+ and LFCN-1200+
	high and low pass filter, respectively, as well as two TCD-13-4X
	couplers used to couple off a copy of the input signal and to couple
	in a low-frequency calibration tone. The bandpass of the input chain
	is primarily determined by the high and low pass filters, which act to
	block DC and as an antialiasing filter for the digitizer. A
	high-frequency impulse generated by a Tektronix AWG5104 was digitized by
	both the SURFv5 and a Tektronix MSO 5204B oscilloscope with a \SI{2}{\giga\hertz} input bandwidth under
	identical conditions. Waveforms from the SURFv5 were correlated and
	averaged to produce the upsampled waveform, shown in Figure~\ref{fig:pulse_vs_windowed}, along
	with the impulse viewed by the oscilloscope, shown in Figure~\ref{fig:AVG_waveform} for comparison.
	
\begin{figure}[t]
		\centering
		\begin{minipage}[t]{0.48\linewidth}\centerline{\includegraphics[height=6cm]{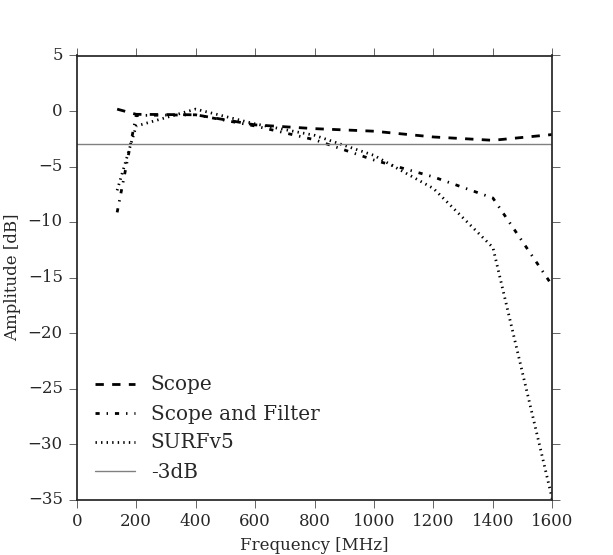}}
			\caption{The response of the SURFv5, the SURFv5 plus the bandpass filter, and the oscilloscope.}
			\label{fig:BW_vs_scope_bw}
		\end{minipage}
		\hspace{0.02\linewidth}
		\begin{minipage}[t]{0.48\linewidth}\centerline{\includegraphics[height=6cm]{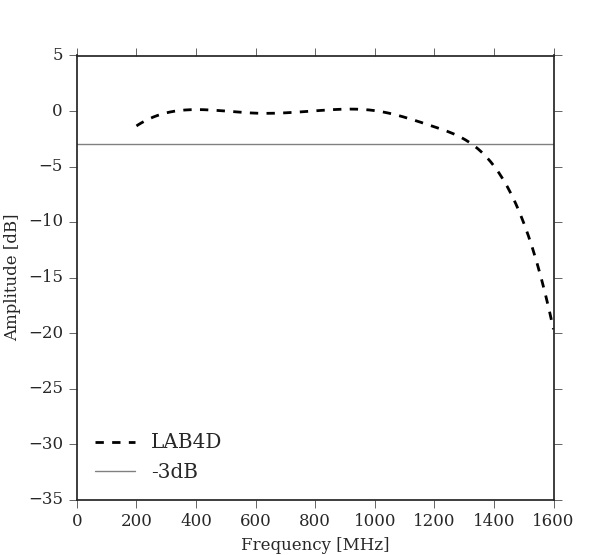}}
			\caption{LAB4D de-embedded bandwidth measurement.}
			\label{fig:BW}
		\end{minipage}
	\end{figure}

	To extract the small-signal analog bandwidth, the Fourier transform of
	the impulse response for both the SURFv5 and the oscilloscope were
	taken using a $\pm$\SI{10}{\nano\s} Hann window around the peak to eliminate
	effects from reflections due to imperfect input matching at the
	coaxial cable connections. The SURFv5 input bandpass was then obtained
	from a test board using a network analyzer and applied to the
	oscilloscope response. Finally, the LAB4D de-embedded frequency
	response was obtained by subtracting the measured SURFv5 response from
	the modified oscilloscope response. The oscilloscope and the de-embedded LAB4D
	response are shown in Figure~\ref{fig:BW_vs_scope_bw}.

	The measured -3\hspace{0.05cm}dB point of the LAB4D is approximately $1.3\,\textrm{GHz}$, with a
	frequency response flat to within $0.5\,\textrm{dB}$ observed up to
	$1.1\,\textrm{GHz}$. This is a conservative measurement, as
	uncertainty in correlating the corresponding waveforms from the SURFv5
	when averaging them will reduce the high-frequency response. The low-frequency
	response is purely determined by the high-pass filter on the SURFv5. The LAB4D
	has no intrinsic low-frequency limitation and the frequency response below
	\SI{200}{\mega\hertz} is expected to be flat. This represents a 44\% increase over
	the LAB3, which had a measured -3~dB point of \SI{900}{\mega\hertz}.

	\section{Summary}

	A switched capacitor array device developed in the TSMC \SI{0.25}{\micro\m} CMOS (LO) process which utilizes a unique ``ping-pong" intermediate storage array architecture has been designed, fabricated, and characterized. This ASIC, the LAB4D, has a -3\hspace{0.05cm}dB upper analog bandwidth limit of greater than \SI{1.3}{\giga\hertz}, a sampling frequency of 3.2\hspace{0.05cm}GSa/s (well above Nyquist minimum for the 200-1200MHz ANITA band), a sample window length of \SI{320}{\nano\s}, and features the unique ability to tune the timebase sampling offsets, reducing the RMS time variance between sample cells to be reliably less than \SI{5}{\pico\s}.


	\section{Acknowledgments}
	
	This work is the result of the continued support from NASA, the National Science Foundation, and the US Dept. of Energy, High Energy Physics Division.


	
	
	\bibliography{references}


\end{document}